# Feedback suppression in 405 nm superluminescent diodes via engineered scattering

ANDREA MARTÍNEZ PACHECO[1,§], ANTONIO CONSOLI[1,2], CEFE LÓPEZ[1]

[1]*Instituto de Ciencia de Materiales de Madrid (ICMM), Consejo Superior de Investigaciones Científicas (CSIC), Calle Sor Juana Inés de la Cruz, 3, 28049 Madrid, Spain.*
[2]*Escuela de Ingeniería de Fuenlabrada (EIF), Universidad Rey Juan Carlos (URJC), Camino del Molino 5, 28942 Fuenlabrada, Madrid, Spain.*
§*andrea.martinez@csic.es*

Superluminescent diodes are promising devices for applications in which low coherence, high efficiency, small footprint and good optoelectronic integration are required. Blue emitting superluminescent diodes with good performances and easy fabrication process are sought for next generation solid state lighting devices, micro-projectors and displays. These devices are laser diodes in which the optical feedback is inhibited, and lasing action avoided. Conventional fabrication processes minimize optical feedback by *ad-hoc* designs, e.g. anti-reflection coating, tilted waveguide or absorber sections, requiring specific fabrication steps. In this work, we propose and demonstrate the introduction of scattering defects in the device's waveguide as a method for feedback inhibition. By performing pulsed laser ablation on a commercial 405 nm GaN laser diode we demonstrate a superluminescent diode, featuring a maximum output power of 2 mW and a spectral width of 5.7 nm.

## 1. INTRODUCTION

Superluminescent diodes (SLDs) are electrically pumped semiconductor devices in which light emission is produced by the amplified spontaneous emission (ASE) obtained along a waveguide [1,2]. For their purpose, SLDs pursue intense emission without lasing and, therefore, at variance with laser diodes (LDs), optical feedback must be avoided which is achieved with a malformed optical cavity. The lack of a proper optical cavity results in broader emission spectrum and lower temporal and spatial coherence, compared to LDs. This makes SLDs interesting for applications in which speckle artifacts inherent to LDs are detrimental while high energy efficiency is desired as, for example, in pico-projectors [3], displays [4], optical coherence tomography (OCT) [5], visible light communications [6,7] and fiber optic gyroscopes [8].

Short wavelength of emission in the visible range is of benefit for these applications as it allows white light emission required in lighting and projection, and high spatial resolution in OCT [9]. Blue SLDs based on gallium nitride (GaN) were firstly demonstrated in 2009 utilizing angled waveguides [10] and roughened facets [11]. Since then, research in the field has been continuously growing [12,13]. At 405 nm emission wavelength, typical performances in terms of spectral width and output power and are in the order of few nanometers and few to tens of milliwatts [14–16], also in commercially available devices.

Only recently, a spectral width of 12.5 nm and a maximum output power of 15 mW have been achieved, by combining highly tilted facet angle and an absorber section [7].

Structurally, SLD architecture recalls edge emitting LDs, however, the design and fabrication process of SLDs are intended to suppress optical feedback and avoid lasing action, while still providing ASE. Four different approaches have been reported: i) anti-reflection coatings of cavity facets, ii) bending or tilting the waveguide with respect to the output facet, iii) adding an absorber section and iv) roughening the facet surface [1].

Applying anti-reflection coatings to the facets of a LD is the simplest approach to minimize feedback, however, very low reflection coefficients (in order of $10^{-5}$) are required, which is challenging in real structures [17], limiting anti-reflection coatings as a complementary method combined with others.

Tilting or bending the waveguide relative to the output facet results in an inclined output beam and requires a precise calculation of parameters such as the tilt angle, effective refractive index and ridge width [18,19].

Integrating an absorber section into the waveguide introduces additional losses [20] but is unsuitable for high injection currents due to loss compensation and onset of lasing action.

Use of scattering defects to disrupt the cavity has been poorly investigated, see [11], where GaN SLDs have been obtained by wet etching of one facet of a LD.

In this work, we describe a novel fabrication approach in which scattering defects are introduced into the optical waveguide of a commercially available GaN LD, avoiding optical feedback and lasing action. The mirrors of the original LD are kept untouched, maintaining a straight un-tilted output beam with low divergence angles and straight emission that could favor applications relying on fiber coupling for instance. Scattering defects are obtained via pulsed laser ablation of the optical waveguide, by milling a hole into the ridge waveguide located on the top of the device. We characterized the emission properties of the modified LD, obtaining broad emission spectrum, good directionality and low spatial coherence. In our vision, the value of the presented research is the demonstration of a novel and simple fabrication process of 405 nm SLDs based on pulsed laser ablation of commercial LDs which results in good performances for low coherence applications.

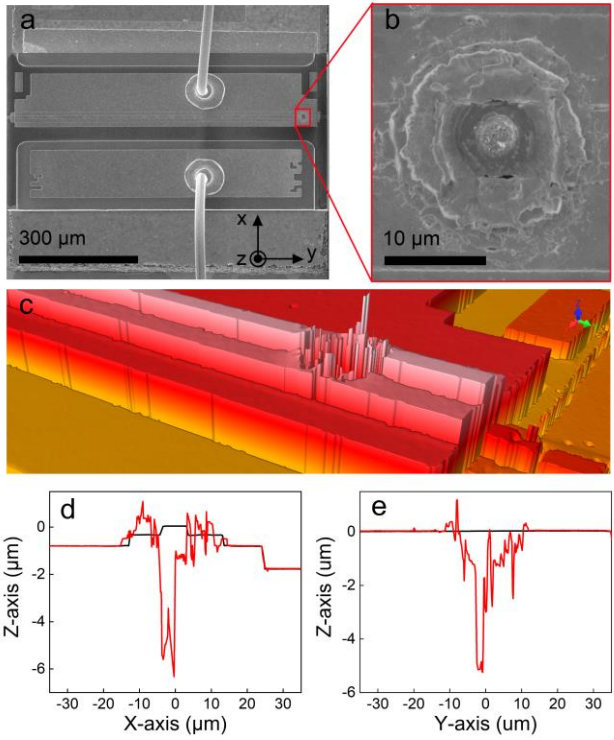

**Fig. 1**. SEM images (a, b) and profiles of the modified LD (c-e). Modified device seen from above (a): emission occurs in the plane of the image and towards the left. The ridge waveguide is in the upper patch whereas the lower patch corresponds to the bottom electrode. Blow up of the ablated hole (b). 3D profile of the ablated region (c). Cross section of the ablated area along the x-axis (d) and y-axis (e) for the original (black lines) and modified (red lines) devices.

## 2. EXPERIMENTAL RESULTS

### A. Device modification

The original device is a commercially available GaN LD emitting at 405 nm (Ushio HL40053MG). The device is delivered in a TO-can packaging which is opened for device inspection and modification. The LD threshold current is 90 mA and maximum output power of 300 mW at 400 mA.

The LD is mounted on a metallic L-shaped holder fabricated ad-hoc which allows imaging the top surface of the dice, observing the FP cavity from above. The modification procedure consisted of controlled pulsed laser ablation of the ridge waveguide which is accessible from above the device. The ablation process is performed by using a mode-locked, frequency-doubled (532 nm) Nd:YAG laser (Ekspla model, 10 Hz repetition rate, 30 ps pulse duration). The pulsed laser delivered 2 pulses with pulse energy of 80 µJ, and diameter 18 µm, located at about 50 µm from the back mirror, see Fig 1. In Fig. 1a, a top view of the modified LD is provided. The LD has a narrow (about 8 µm) ridge waveguide (dim, narrow, long, rectangle at the bottom of the top electrode patch) and a Fabry-Perot (FP) cavity length of 790 µm.

The ablation process bored a hole into the metallic contact on top of the ridge and dug a pit into the ridge waveguide, with rough walls and irregular features, see Fig. 1b. The 3D profile of the ablated region has been obtained by optical profilometry (Fig. 1c), and the transverse (x-axis) and longitudinal (y-axis) cross sections are shown in Fig. 1d and 1e, respectively. The maximum depth of the hole is about 5 µm and it is characterized by rough internal walls with irregular profiles. Transverse and longitudinal profiles show that the ablation depth reached far below the active layer which consists of multiple quantum wells and is located below the ridge waveguide. Light scattered from the hole is observed when the device is biased with few tens of mA.

### B. Emission characteristics

After ablation, the modified LD is mounted on a temperature and current controller (Arroyo Instrument 6305) connected to a PC, for emission characterization. The device case is kept at 20º C and the power emitted from the output mirror is measured with a calibrated photodiode (Thorlabs S130C). The maximum output power is 2 mW at 350 mA, as shown in Fig. 2a.

The output power curve observed in Fig. 2a is typical of strongly injected SLDs [6,10] with linear dependence for high current values. This is attributed to a non-linear dependence of gain [21] on current and can be modeled as [10,11,22]: $P(I) \propto (\Gamma \cdot g(I) - \alpha)^{-1} \cdot [\exp(\Gamma \cdot g(I) - \alpha) \cdot L]$, where $\Gamma$ is the confinement factor, $g(I)$ is the current dependent gain, $\alpha$ are internal losses and $L$ is the cavity length.

We introduced a simple expression of the gain $g(I)$, which is defined as $g(I) = 1 - \exp(g_0 \cdot I)$, and we considered $g_0$, $\Gamma$ and $\alpha$ as parameters in the numerical fitting procedure. Simulation results are plotted in Fig. 2a. The linear region of output power is due to the decrease of the differential gain, *i.e.*, the local slope of the gain curve shown in Fig. 2a, at high currents [21].

Spectral measurements are performed by collimating the output emission with a microscope objective (×100, 0.9 NA) and focusing into an optical fiber (diameter 100 µm) connected to a spectrometer (Andor Technology, 313i-A), with spectral resolution of 0.15 nm.

The central wavelength of emission blue-shifts from 406.65 nm to 402.15 nm when current is varied between 20 mA to 350 mA, see Fig. 2b. This is typical of SLDs [6,14,20] and is attributed to the band filling effect caused by increasing carrier density [23], as in the ASE regime and in LDs below threshold, where the carrier density is not clamped to the threshold value. It is opposite to the red shift in LDs above threshold due to constant carrier density and Joule heating. The full width at half maximum (FWHM) of the emission spectra as a function of current is also shown in Fig. 2b, observing a spectral narrowing from 12 nm to 5.7 nm.

In the original LD, the output power and the spectral width at maximum injected current corresponded to 400 mW and 0.6 nm, respectively. The observed decrease of power and increase of spectral width is attributed to the losses added by the modification process by pulsed laser ablation, as observed in random laser diodes we previously fabricated with similar process [24].

Emitted spectra are shown in Fig. 2c for different bias currents. The spectral profiles consist of bell-shaped curves with few nanometers width and superimposed narrow peaks with sub-nanometer widths, not to be mistaken for noise. This is understood as ASE with residual modes from the disordered cavity formed after the ablation process. The proposed device is actually a disordered cavity consisting of the output mirror and the scattering walls of the ablated hole, working in ASE regime. In fact, any SLD could be driven into lasing emission, provided it could withstand the current producing enough gain for even small back reflections into the active medium to close a feedback loop [16]. In an ordinary FP cavity, the residual spectrum modulation is periodic, while in disordered cavities peaks are randomly distributed in frequency and mode competition makes them vary with current [25], as observed in Fig. 2c.

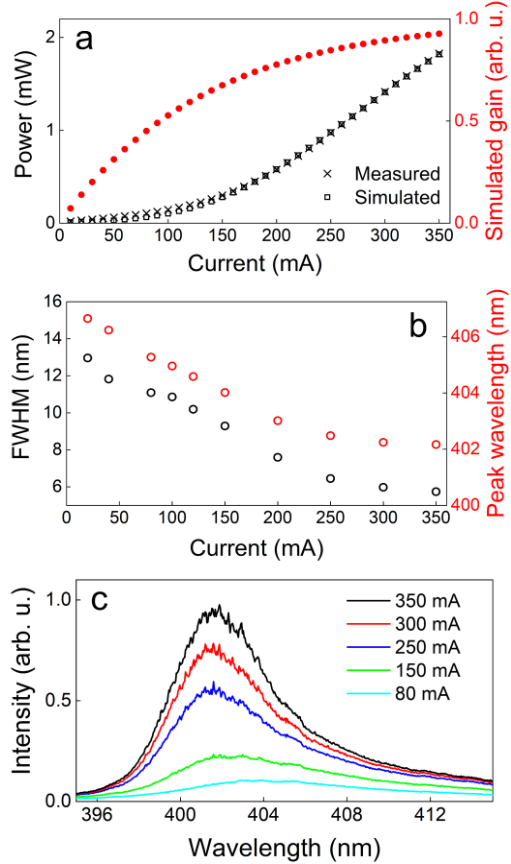

**Fig. 2.** (a) Comparison of the intensity measured (cross) and simulated (circles) versus current with the corresponding simulated gain of the modified diode. (b) Peak wavelength and FWHM versus current. (c) Emitted spectra at different injection currents.

The emission profiles of the output beam of the original LD and the fabricated SLD, collimated with a lens placed at its focal length from the SLD facet, and recorded with a CMOS camera are shown in Fig. 3a and 3b, respectively. From both devices we observe an elliptical beam which is larger for the SLD compared to the original LD. In Fig. 3c, we plot the beam profiles of the SLD along the parallel and orthogonal directions to the active region. The divergence angles, measured at $1/e^2$, are 11.6 ° and 34.5 ° along the parallel and perpendicular directions, respectively. This corresponds to a small increase with respect to the original LD, which shows divergence angles of 6.2 ° and 31.6 ° along the parallel and perpendicular directions, respectively. Since the divergence parallel to the active region plane is governed by the width of the emitting area, which is many wavelengths in size, it does not change much upon processing (10% increase). However, the spread in the perpendicular plane is heavily affected by processing (twofold increase).

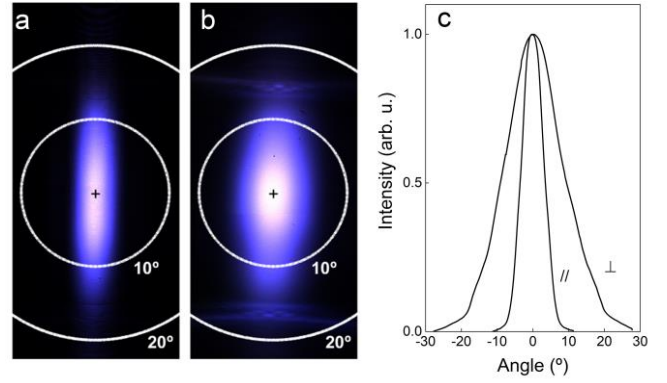

**Fig. 3.** Emission profiles of the output beam of the original LD (a) and modified SLD (b). Parallel and perpendicular cross sections of directions of the SLD beam (c).

The output beam is also characterized by measuring the speckled contrast produced by shining a diffusing surface with the collimated output beam and imaging the resulting pattern onto a CMOS camera [26]. Speckle images of the original and modified devices when biased at 300 mA are shown in Fig. 4a and 4b, respectively. Largely dispersed intensity values and higher contrast are observed from the original LD due to the constructive/disruptive interference at the image plane. More uniform intensity values are obtained from the SLD, resulting in reduced speckle contrast due to lower spatial coherence.

The probability density function, $P(I)$, normalized to the average intensity value, is plotted in Fig. 4c, for the original LD (black dots) and the SLD (blue dots), when both biased with 300 mA. The speckle contrast, $C$, is calculated as $C = \sigma_I / \langle I \rangle$, where $\sigma_I$ and $\langle I \rangle$ are the standard deviation and the mean of the detected pixel intensities, respectively. The contract $C$ gives direct information regarding the number of modes creating the pattern, $N = C^{-2}$ [26]. A narrower distribution is observed for the SLD, indicating a lower degree of spatial coherence, with $C < 0.2$, indicating a larger number of modes $N$, above 25. In Fig. 4d, the speckle contrast as a function of current is shown for the original LD and the SLD. We observe similar values for both devices below 100 mA (the LD threshold is of 90 mA), a sudden increase around threshold for the LD and a smooth increase for the SLD with current. For all values, $C$ is greater in LD than in the SLD, this is understood as a lower degree of spatial coherence in the fabricated device.

## 3. CONCLUSIONS

A new method for the fabrication of GaN SLD is proposed and demonstrated. We performed pulsed laser ablation of the ridge waveguide of an existing GaN LD so that the feedback loop in the cavity is broken by the introduction of scattering defects into the light amplification path of the original LD. We have characterized the fabricated SLD, obtaining a maximum output power of 2 mW, a spectral width of 5.7 nm, good directionality and low spatial coherence.

We demonstrated the use of scattering elements as a practical tool for feedback and lasing suppression in optical cavities, which has been scarcely reported to date. We consider that our approach would help the development of novel techniques for SLD fabrication, due to its simple implementation

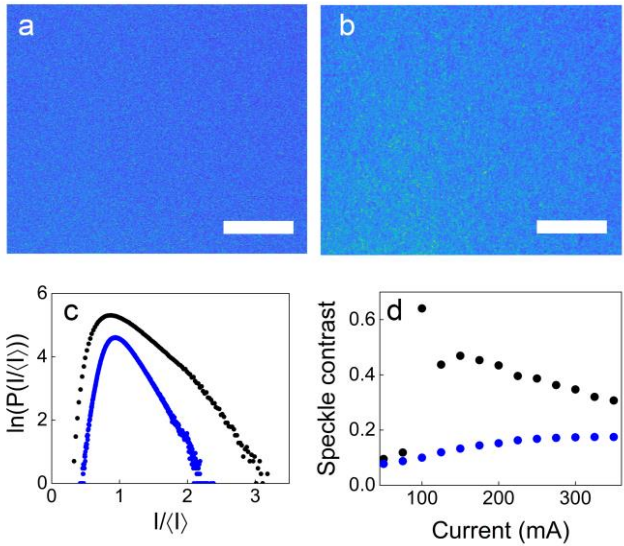

**Fig. 4.** Speckle images of the original (a) and modified (b) devices when biased at 300 mA. Probability density function (c) for the original LD (black dots) and the SLD (blue dots), for an injection current of 300 mA. Speckle contrast (d) as a function of current for the original LD (black dots) and the SLD (blue dots).

and the unexplored potential of optical cavities with scattering defects.

The added value of our method is its simplicity, as it can be performed after completion of the fabrication process of a commercial LD and does not require the adoption of standard fabrication techniques, e. g. realization of bent or tilted waveguides, anti-reflection mirror coating or addition of a passive section into the cavity, which result in high realization costs.

## Funding

This work has received funding from Ministerio de Ciencia, Innovación y Universidades through the procets RanLaD (PDC2022-133418-I00) and SPhAM (PID2021-124814NB-C21).

## Disclosure

The authors declare no conflicts of interest.

## Data availability.

The data that support the plots within this paper are available from the corresponding authors upon reasonable request.

## REFERENCES

1. H. Lu, O. Alkhazragi, Y. Wang, N. Almaymoni, W. Yan, W. H. Gunawan, H. Lin, T.-Y. Park, T. K. Ng, and B. S. Ooi, "Low-coherence semiconductor light sources: devices and applications," npj Nanophotonics **1**, 9 (2024).
2. S. Stanczyk, A. Kafar, D. Schiavon, S. Najda, T. Slight, and P. Perlin, "Edge Emitting Laser Diodes and Superluminescent Diodes," in *Nitride Semiconductor Technology* (2020).
3. F. Kopp, C. Eichler, A. Lell, S. Tautz, J. Ristić, B. Stojetz, C. Hö, T. Weig, U. T. Schwarz, and U. Strauss, "Blue superluminescent light-emitting diodes with output power above 100mW for picoprojection," Jpn. J. Appl. Phys. **52**, (2013).
4. M. Rossetti, A. Castiglia, M. Malinverni, C. Mounir, N. Matuschek, M. Duelk, and C. Vélez, "RGB Superluminescent Diodes for AR Micro-Displays," in *Digest of Technical Papers - SID International Symposium* (2018), Vol. 49.
5. R. A. Costa, M. Skaf, L. A. S. Melo, D. Calucci, J. A. Cardillo, J. C. Castro, D. Huang, and M. Wojtkowski, "Retinal assessment using optical coherence tomography," Prog. Retin. Eye Res. **25**, (2006).
6. A. A. Alatawi, J. A. Holguin-Lerma, C. H. Kang, C. Shen, R. C. Subedi, A. M. Albadri, A. Y. Alyamani, T. K. Ng, and B. S. Ooi, "High-power blue superluminescent diode for high CRI lighting and high-speed visible light communication," Opt. Express **26**, (2018).
7. M. Dong, S. Yi, J. Wang, C. Ma, D. Li, S. Wang, Y. Hou, Z. Li, J. Zhang, J. Shi, N. Chi, and C. Shen, "High-Speed GaN-Based 405 nm Violet Superluminescent Diode with Tilted Facet for Visible Light Communications," Phys. Status Solidi Appl. Mater. Sci. **221**, (2024).
8. J. Nayak, "Fiber-optic gyroscopes: From design to production [Invited]," Appl. Opt. **50**, (2011).
9. G. R. Goldberg, A. Boldin, S. M. L. Andersson, P. Ivanov, N. Ozaki, R. J. E. Taylor, D. T. D. Childs, K. M. Groom, K. L. Kennedy, and R. A. Hogg, "Gallium Nitride Superluminescent Light Emitting Diodes for Optical Coherence Tomography Applications," IEEE J. Sel. Top. Quantum Electron. **23**, (2017).
10. E. Feltin, A. Castiglia, G. Cosendey, L. Sulmoni, J. F. Carlin, N. Grandjean, M. Rossetti, J. Dorsaz, V. Laino, M. Duelk, and C. Velez, "Broadband blue superluminescent light-emitting diodes based on GaN," Appl. Phys. Lett. **95**, 16–18 (2009).
11. M. T. Hardy, K. M. Kelchner, Y. Da Lin, P. S. Hsu, K. Fujito, H. Ohta, J. S. Speck, S. Nakamura, and S. P. DenBaars, "m-plane GaN-based blue superluminescent diodes fabricated using selective chemical wet etching," Appl. Phys. Express **2**, 1–3 (2009).
12. A. Kafar, S. Stanczyk, D. Schiavon, T. Suski, and P. Perlin, "Review—Review on Optimization and Current Status of (Al,In)GaN Superluminescent Diodes," ECS J. Solid State Sci. Technol. **9**, (2020).
13. N. Feng, M. Lu, S. Sun, A. Liu, X. Chai, X. Bai, J. Hu, and Y. Zhang, "Light-Emitting Device Based on Amplified Spontaneous Emission," Laser Photonics Rev. **17**, (2023).
14. K. Holc, Ł. Marona, R. Czernecki, M. Boćkowski, T. Suski, S. Najda, and P. Perlin, "Temperature dependence of superluminescence in InGaN-based superluminescent light emitting diode structures," J. Appl. Phys. **108**, (2010).
15. C. Shen, C. Lee, T. K. Ng, S. Nakamura, J. S. Speck, S. P. DenBaars, A. Y. Alyamani, M. M. El-Desouki, and B. S. Ooi, "High-speed 405-nm superluminescent diode (SLD) with 807-MHz modulation bandwidth," Opt. Express **24**, (2016).
16. A. Kafar, S. Stańczyk, S. Grzanka, R. Czernecki, M. Leszczyński, T. Suski, and P. Perlin, "Cavity suppression in nitride based superluminescent diodes," J. Appl. Phys. **111**, (2012).
17. D. Marcuse, "Reflection Loss of Laser Mode from Tilted End Mirror," J. Light. Technol. **7**, (1989).
18. G. A. Alphonse, "<title>Design of high-power superluminescent diodes with low spectral modulation</title>," in *Test and Measurement Applications of Optoelectronic Devices* (2002), Vol. 4648.
19. A. Kafar, S. Stańczyk, P. Wiśniewski, T. Oto, I. Makarowa, G. Targowski, T. Suski, and P. Perlin, "Design and optimization of InGaN superluminescent diodes," Phys. Status Solidi Appl. Mater. Sci. **212**, (2015).
20. C. Shen, T. K. Ng, J. T. Leonard, A. Pourhashemi, S. Nakamura, S. P. DenBaars, J. S. Speck, A. Y. Alyamani, M. M. El-desouki, and B. S. Ooi, "High-brightness semipolar (2021⁻) blue InGaN/GaN superluminescent diodes for droop-free solid-state lighting and visible-light communications," Opt. Lett. **41**, (2016).
21. A. Kafar, S. Stanczyk, G. Targowski, T. Suski, and P. Perlin, "Gain saturation in InGaN superluminescent diodes," Gall. Nitride Mater. Devices IX **8986**, 89860P (2014).
22. R. Cahill, P. P. Maaskant, M. Akhter, and B. Corbett, "High power surface emitting InGaN superluminescent light-emitting diodes," Appl. Phys. Lett. **115**, (2019).
23. "Fundamentals of semiconductor lasers," Springer Ser. Opt. Sci. **93**, (2014).
24. A. Consoli, N. Caselli, and C. López, "Electrically driven random lasing from a modified Fabry–Pérot laser diode," Nat. Photonics **16**, 219–225 (2022).
25. A. Consoli, P. D. García, and C. López, "Tuning the emission properties

of electrically pumped semiconductor random lasers via controlled pulsed laser ablation," Opt. Express **31**, 42439 (2023).
26. M. Nixon, B. Redding, A. A. Friesem, H. Cao, and N. Davidson, "Efficient method for controlling the spatial coherence of a laser," Opt. Lett. **38**, (2013).
27. J. W. Goodman, "Some fundamental properties of speckle*," J. Opt. Soc. Am. **66**, 1145 (1976).